\documentstyle[prd,preprint,floats,epsfig,aps,12pt]{revtex}
\tightenlines

\begin{document}
\preprint{CFNUL/98-04}
\draft
\title{Supernova neutrino oscillations:\\
Adiabaticity improvement by Majoron fields}
\author{Lu\'{\i }s Bento}
\address{Centro de F\'{\i }sica Nuclear da Universidade de Lisboa,\\
Avenida Prof. Gama Pinto 2, 1699 Lisboa - {\sl codex}, Portugal\\
{\rm e-mail: lbento@fc.ul.pt} \\
\strut }
\date{hep-ph/9806411, 7 June 1998, last change 8 November 1998}
\maketitle
\begin{abstract}
If the lepton numbers are associated with global symmetries spontaneously
broken at a scale below 1 TeV, neutrino oscillations in supernovae produce
classic Majoron fields that perturb the neutrino propagation itself and may
change the oscillation patterns in the periods of largest $\nu $ fluxes. The
impact of the Majoron fields on the same transitions as $\nu _{e}\rightarrow
\nu _{X}$ that presumably occur in the Sun is studied in the case of the
non-adiabatic MSW solution. It is shown how the back reaction of the Majoron
fields may improve the adiabaticity of these oscillations in a supernova
environment which has implications on the outgoing $\nu _{e}$ spectrum.
\end{abstract}
\strut





\section{Introduction}

Several experiments from solar to atmospheric neutrinos and
laboratory oscillation experiments \cite{cald98} indicate that neutrinos
oscillate and leptonic flavors are not conserved. This is in contrast
with the minimal standard model (SM). It is quite possible that the energy
scale of breaking of the three lepton numbers is comparable or even smaller
than the Fermi scale. Furthermore, if they are spontaneously broken by the
expectation values of some scalar fields then, a few bosons with zero mass
should exist - the Nambu-Goldstone bosons (NG) - one per broken global
symmetry.

It was recently pointed out\thinspace \cite{bent98} that these NG bosons
(called Majorons or familons when associated with lepton numbers) couple to
the time rate of creation of the respective lepton numbers carried by the
matter particles and therefore coherent NG fields are produced whenever
lepton number violating processes occur simultaneously. That is the case if
neutrinos change flavor on their way out from stars as seems to happen in
the solar system. Once the NG fields are generated (the triggering process
may be a normal Mikheyev-Smirnov-Wolfenstein (MSW) resonant conversion \cite
{mikh86}), they change in turn the relative potentials of the different
neutrino species and so get a life of their own.

\strut The numbers show that if the scale of symmetry breaking is below 1
TeV then, the Majoron fields are important enough to play a role in
supernova neutrino oscillations. They are however too small in the case of
the Sun unless the scale of symmetry breaking lies below 1 KeV. As a result,
supernova neutrinos may exhibit oscillation patterns in contradiction with
the observations of solar, atmospheric and terrestrial neutrinos. We ought
to be prepared, in the event of a close by supernova explosion, for the
possible kind of effects caused by NG fields.

In the previous paper \cite{bent98}, the example studied was that Majoron
fields generated by the conversion $\nu _{e}\rightarrow \nu _{\tau }$,
assumed to take place in a certain resonance shell, yield neutrino
potentials which become competitive with the standard electroweak potentials
at larger radii and therefore affect the other flavor transitions
characterized by smaller $\Delta m^{2}$. The effects can be so dramatic as
the resonant oscillation $\bar{\nu}_{e}\leftrightarrow \bar{\nu}_{\mu }$ in
a context of $m_{\nu_e}<m_{\nu_{\mu}}$ hierarchy, where the resonance is otherwise
possible for $\nu_{e} \leftrightarrow \nu_{\mu }$ but not for the
anti-neutrinos, if they only interact via standard $W$ and $Z$ bosons. In
the present paper, I want to discuss what happens if the Majoron potentials
are already significant in the very region where the oscillations they are
generated from occur. Then, a back reaction effect takes place yielding an
interesting flavor dynamics.

\strut \strut Suppose that in the Sun the electron-neutrino oscillates into
the muon-neutrino with the parameters of the non-adiabatic, small mixing
angle solution \cite{haxt86,rose86}. Then, the $\nu _{e}\leftrightarrow \nu
_{\mu }$ transitions are also non-adiabatic in a supernova and only a
fraction of each neutrino species is converted into the other. Furthermore,
since the level crossing probability is an increasing function of the
energy, the hotter $\nu _{\mu }$s\ have larger survival probabilities than
the cooler $\nu _{e}$s. It will be shown that the back reaction of the NG
fields improves the adiabaticity of the neutrino transitions, thus yielding
a hotter energy spectrum for the outgoing $\nu _{e}$s. That kind of effect
can in principle be traced in those detectors such as Super-Kamiokande and
SNO, capable of detecting supernova electron-neutrinos \cite{burr92}.

\section{Majoron and neutrino flavor dynamics}

\strut \strut

\strut In the following it is assumed that the partial lepton number $L_{e}$
is conserved at the Lagrangian level but the global symmetry associated with
it, U(1)$_{e}$, is spontaneously broken by the expectation values of one or
more scalar iso-singlets $\sigma _{i}$. Then, a NG boson $\xi _{e}$ exists
with zero mass. The neutrino mass matrix violates in principle the three
lepton numbers but for simplicity I will ignore the other possibly existing
Majorons. It may be interpreted as meaning that the respective scales of
symmetry breaking are slightly higher.

It is well established that the Nambu-Goldstone bosons only interact through
derivative couplings \cite{chen89,wein96}\ (related to the soft pion low
energy theorems). This is clarified \cite{gelm83,bent98} by changing
variables from the original fields with definite $L_{e}$ charges namely,
fermions $\chi ^{a}$ and scalars $\sigma _{i}$, as follows:

\begin{eqnarray}
\chi ^{a} &=&\exp (-i\xi _{e}L_{e}^{a})\,\psi ^{a}\ ,  \label{psi} \\
&&  \nonumber \\
\sigma _{i} &=&\exp (-i\xi _{e}L_{e}^{i})\,\left( \left\langle \sigma
_{i}\right\rangle +\rho _{i}\right) \ .  \label{sigma}
\end{eqnarray}
The so defined {\em physical} weak eigenstates, the fermions $\psi
^{a}=e,\,\nu _{e},\,...$\ and massive bosons $\rho _{i}$, are invariant
under the group U(1)$_{e}$. In terms of these fields the symmetry is simply
realized as translations\ of the Majoron field, $\xi _{e}\rightarrow \xi
_{e}+\alpha $, and consequently the Lagrangian can only depend on the
derivatives $\partial _{\mu }\xi _{e}$.

\strut The non-standard interactions relevant to this work are contained in
the expression

\begin{equation}
{\cal L}=-{{\nu _{L}^{a}}^{T}\,}C\,m{_{ab}\,\nu _{L}^{b}\,}+{\rm h.c.}+\frac{%
1}{2}\Omega _{e}^{2}\,\partial _{\mu }\xi _{e}\,\partial ^{\mu }\xi _{e}+\,%
{\cal L}_{{\rm int}}(\partial _{\mu }\xi _{e})\ ,  \label{LbSM}
\end{equation}
where $\Omega _{e}^{2}\,=2\sum_{i}\left| L_{e}^{i}\,\left\langle \sigma
_{i}\right\rangle \right| ^{2}$ and $\nu _{L}^{a}$ denote the standard $\nu
_{L}^{e},\nu _{L}^{\mu },\nu _{L}^{\tau }$ or any extra neutrino singlets.
The $\xi _{e}$ equation of motion, 
\begin{equation}
\partial _{\mu }\partial ^{\mu }\,\xi _{e}=-\partial _{\mu }J_{e}^{\mu
}/\Omega _{e}^{2}\ ,  \label{ddfi}
\end{equation}
identifies with the conservation law of the N\"{o}ether current associated
with the symmetry $\xi _{e}\rightarrow \xi _{e}+\alpha $. All the Majoron
interactions are cast in the current $J_{e}^{\mu }$ obtained from ${\cal L}_{%
{\rm int}}(\partial _{\mu }\xi _{e})$. Its leading terms do not depend on
the particular model and are derived from the $\chi ^{a}$, $\sigma _{i}$
kinetic Lagrangian using Eqs.\ (\ref{psi}), (\ref{sigma}). The result is the
following:
\begin{mathletters}
\begin{eqnarray}
{\cal L}_{{\rm int}} &=&(\partial _{\mu }\xi _{e})\, \left( {\bar{e}\,}\gamma
^{\mu }\,e+{\bar{\nu}_{e}}\gamma ^{\mu }\nu _{e}+\cdots \right) \ ,
\label{Lint} \\
  &  &  \nonumber \\
J_{e}^{\mu } &=&\,{\bar{e}\,}\gamma ^{\mu }\,e+{\bar{\nu}_{e}}\gamma ^{\mu
}\nu _{e}+\cdots \ .  \label{Je}
\end{eqnarray}
The dots represent scalar boson terms and model dependent radiative
corrections that are not relevant for this work and will ignored in the
following.

\strut The expressions above are typical of one kind of models, the
Abelian-singlet Majorons \cite{chik81,vall82,grin85,bent98}. Singlet means
that the scalar fields that spontaneously break the lepton number symmetries
are all singlets under the SM gauge group SU(2)$\times $U(1), thus complying with the LEP results on the $Z^{0}$ invisible width, unlike the
triplet-Majoron model \cite{gelm81}. By Abelian I mean U(1) symmetry groups,
not horizontal in flavor space, associated to $L_{e},$ $L_{\mu
},$ $L_{\tau }$ or any linear combinations of them. As the respective
currents do not change flavor these models are not bound by the laboratory
limits on the familon models \cite{pdg96}. In addition, of all interactions
involving SM particles the neutrino masses are the most important source of
lepton numbers violation. Thus, in single collision or decay reactions, the
effective strength of the Majoron couplings, resulting from $\partial _{\mu
}J^{\mu }/\left\langle \sigma \right\rangle ,$ is proportional to the
neutrino masses and suppressed by the symmetry breaking scale: $g\sim m_{\nu
}/\left\langle \sigma \right\rangle $. For that reason, one does not expect
observable zero neutrino double beta decays accompanied by Majoron emission
as shown in \cite{hirs96}. With a sensitivity to neutrino-Majoron couplings\ of
the order of $10^{-5}$ \cite{bern92} they cannot even probe relatively low
symetry breaking scales. This kind of models also escape present astrophysical
bounds on the couplings of Nambu-Goldstone\ bosons. Neutrino-Majoron
couplings\ with strengths $g\sim m_{\nu }/\left\langle \sigma \right\rangle $
\ are too far from $10^{-4}$ to change supernova collapse dynamics \cite
{kolb82}, and even below the threshold of $\sim 10^{-8.5}$ for supernova
cooling through singlet Majoron emission \cite{choi90}. Finally, the
pseudo-scalar couplings to electrons, that could be responsible for energy
loss in stars \cite{raff97}, only arise through radiative corrections and
are so further suppressed.

In the $\xi _{e}$ equation of motion, the source term $\partial _{\mu
}J_{e}^{\mu }$ is nothing but the time rate of {\em creation} of\ $L_{e}$%
-number carried by {\em matter} particles per unity of volume. If the
neutrinos $\nu _{e}$ and $\nu _{\mu }$ oscillate into each other outside the
supernova neutrinospheres, but not their anti-particles, the net variation
of ${\,}L_{e}$ is given by the difference between the numbers of converted
neutrinos $N(\nu _{\mu }\to \nu _{e})$ and $N(\nu _{e}\to \nu _{\mu })$. In
a stationary regime the fluxes are constant in time and Eq.\ (\ref{ddfi})
reduces to a Poisson equation with a Coulombian solution for $\xi _{e}$ \cite
{bent98}. The gradient $\vec{A}_{e}=-\vec{\nabla}\xi _{e}$ obeys a Gauss
law. In a spherical symmetric configuration it only has a radial component, 
\end{mathletters}
\begin{equation}
A_{e}(r)=-\frac{1}{\Omega _{e}^{2}}\frac{\dot{L}_{e}(r)}{4\pi \,r^{2}}\ ,
\label{Aei}
\end{equation}
that is determined by the integral of the source term over the volume of
radius $r$, in this case, $L_{e}$-number created per unity of time in
that volume, $\dot{L}_{e}(r)=\smallint d^{3}x\ \partial _{\mu }J_{e}^{\mu }$%
. It can also be expressed as \cite{bent98} 
\begin{equation}
A_{e}{\scriptsize (}r{\scriptsize )}=-\frac{1}{\Omega _{e}^{2}}\left[ j({\nu
_{\mu }\to \nu _{e})}-j({\nu _{e}\to \nu _{\mu })}\right] \ ,  \label{Ae}
\end{equation}
where $j({\nu _{e}\to \nu _{\mu })}$ denotes the flux of $e$-neutrinos
converted to $\mu $ flavor and $j({\nu _{\mu }\to \nu _{e})}$ the
reciprocal. Both these quantities are functions of the radius $r$.

\strut Taking in consideration the interactions specified by Eqs.\ (\ref
{LbSM}), (\ref{Lint}), the equation of motion of the neutrino wave function
is, including flavor space, 
\begin{equation}
\left( i\,\partial \hspace{-0.22cm}/\,-\gamma ^{0}V_{0}-\vec{\gamma}\!\cdot
\!\vec{A}_{e}\,\hat{L}_{e}\right) \psi _{L}=m\,\psi _{R}\ ,  \label{empsi}
\end{equation}
where $m$\ is the $\nu $ mass matrix (real for simplicity), $\hat{L}_{e}$\ is
the flavor-valued quantum number (1 for ${\nu _{e}}$ and 0 for ${\nu _{\mu }}
$, ${\nu _{\tau }}$) and $V_{0}$\ designates the flavor conserving SM
potential in a medium at rest. 
One derives in the same way as in the case of a scalar potential $V_{0}$\
the equations governing flavor oscillations \cite{kuo89} (see also \cite{bent98}), 
\begin{equation}
i\frac{\partial }{\partial \,r}\nu =\left( \frac{m^{2}}{2E}+V_{0}+A_{e}\,%
\hat{L}_{e}\right) \nu \ ,
\end{equation}
which give (after absorbing a flavor universal term in the wave function) 
\begin{equation}
i\frac{\partial }{\partial \,r}\left( 
\begin{array}{c}
{\nu _{e}} \\ 
\\ 
{\nu _{\mu }}
\end{array}
\right) =\frac{1}{2E}\left( 
\begin{array}{cc}
2E(V_{W}+A_{e}) & \quad {\rm \,}\frac{1}{2}\Delta m^{2}\sin 2\theta  \\ 
&  \\ 
{\rm \,}\frac{1}{2}\Delta m^{2}\sin 2\theta  & \quad {\rm \,\,}\Delta
m^{2}\cos 2\theta 
\end{array}
\right) \left( 
\begin{array}{c}
{\nu _{e}} \\ 
\\ 
{\nu _{\mu }}
\end{array}
\right) \;.
\end{equation}
$V_{W}=\sqrt{2}\,G_{{\rm F}}\,n_{e}$ is the charged-current potential of $%
\nu _{e}$\ in a medium with electron number density $n_{e}$ \cite{wolf78}\
and $\theta $\ is the mixing angle. It is worth to notice that the results
do not change if one considers alternatively a NG boson associated with
breaking of $L_{\mu }$ or $L_{e}-L_{\mu }$. The reason is, $\nu
_{e}\leftrightarrow \nu _{\mu }$ oscillations only violate $L_{e}-L_{\mu }$
(the non-conservation of $L_{e}+L_{\mu }$ is suppressed by $m_{\nu
}^{2}/E^{2}$) and only care about the difference between the $\nu _{e}$ and $%
\nu _{\mu }$ potentials.

In order to calculate $A_{e}(r)$\ one needs to know the fluxes of $\nu _{\mu
}$ and $\nu _{e}$\ as functions of the radius with and without oscillations.
I make the simplification of neglecting the oscillation length by saying
that a neutrino with energy $E_{R}$ is converted to the other flavor at the
position where the resonance condition, 
\begin{equation}
E_{R}=\frac{\Delta m^{2}\cos 2\theta }{2(V_{W}+A_{e})}\ ,  \label{ER}
\end{equation}
is fulfilled. In addition, having in mind that in a non-adiabatic regime
only a fraction $1-P_{c}$\ changes flavor (small mixing angle), the level
crossing probability $P_{c}$\ is calculated using the Landau-Zener
approximation \cite{haxt86,kuo89},

\begin{equation}
P_{c}(E)=\exp \left\{ -\frac{\pi }{4}\left| \frac{{\rm \,}\Delta m^{2}}{{\rm %
d}E_{R}/{\rm d}r}\frac{\sin ^{2}2\theta }{\cos 2\theta }\right| \right\}
_{\!\!R}\ .  \label{Pc}
\end{equation}
I believe that these approximations change the numbers but not the lesson
taken by comparing the results with and without a Majoron field.

Let the number of emitted particles per unity of time and energy be
specified by the distribution functions $f_{\nu _{e}}(E)={\rm d}\dot{N}_{\nu
_{e}}/{\rm d}E$ and $f_{\nu _{\mu }}(E)={\rm d}\dot{N}_{\nu _{\mu }}/{\rm d}%
E $. They are normalized by the relation between the number luminosity $\dot{%
N}_{\nu }$, the energy luminosity $L_{\nu }$ and the average energy $\bar{E}%
_{\nu }$\ of each $\nu $ flavor namely, $\dot{N}_{\nu }=L_{\nu }/\bar{E}%
_{\nu }$. $L_{\nu }$\ and $\dot{N}_{\nu }$\ will be given below in unities
of ergs/s and ergs/s/MeV respectively. Equation (\ref{ER}) gives the resonance
energy as a function of the radius: particles with lower energies reach the
resonance position at higher density regions and the hottest neutrinos
oscillate at the largest radii. The statement that the number of converted
neutrinos is the fraction $1-P_{c}$\ of the number of particles with
resonance energy translates into an equation for the time rate ${\rm d}\dot{L%
}_{e}={\rm d}\dot{N}({\nu _{\mu }\to \nu _{e})-{\rm d}}\dot{N}({\nu _{e}\to
\nu _{\mu })}$ of $L_{e}$-number creation in a shell with depth ${\rm d}r$: 
\begin{equation}
{\rm d}\dot{L}_{e}=(1-P_{c})\,\left[ f_{\nu _{\mu }}(E_{R})-f_{\nu
_{e}}(E_{R})\right] \frac{{\rm d}E_{R}}{{\rm d}r}{\rm d}r\ .  \label{dLe}
\end{equation}
This establishes a differential equation for $\dot{L}_{e}(r)$. Notice that
the derivative of $E_{R}$\ is not independent from the derivative of $\dot{L}%
_{e}$ because the Majoron potential $A_{e}(r)$\ that enters in $E_{R}$\
depends also on $\dot{L}_{e}(r)$,\ as Eq.\ (\ref{Aei}) shows. In addition,
the probability $P_{c}$\ depends on the $E_{R}$ and $\dot{L}_{e}$
derivatives as well, and that makes a non-linear differential equation for $%
\dot{L}_{e}(r)$ specified by Eqs.\ (\ref{Aei}), (\ref{ER}) - (\ref{dLe}).

\begin{figure}[t]
\centering
\epsfig{file=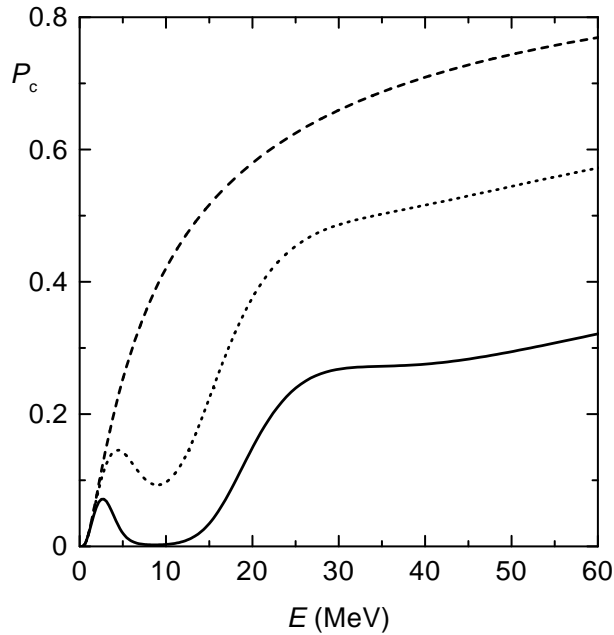,width=80mm}
\vspace{15pt}
\caption{Level crossing probability $P_{c}$ as a function of the $\nu $
energy. Dashed curve holds for the SM with constants $\tilde{M}=4\times
10^{31}\,{\rm g}$\ and $Y_{e}=1/2$. In the dotted and bold curves the
Nambu-Goldstone field $\xi _{e}$ exists with $G_{e}=G_{{\rm F}}$ and $%
G_{e}=4\,G_{{\rm F}}$ respectively, and the luminosities are $10^{52}\,{\rm %
ergs/s}$ for $\nu _{e}$ and $7\times 10^{51}\,{\rm ergs/s}$ for $\nu _{\mu }$%
.}
\end{figure}

It remains to tell the density profile of the medium. In the regions of a
supernova star with densities typical of the Sun the mass density goes as $%
1/r^{3}$, the constant $\tilde{M}=\rho \,r^{3}$ lying between $10^{31}{\rm g}
$ and $15\times 10^{31}{\rm g}$ depending on the star \cite{wils86}. Then,
in terms of the electron abundance $Y_{e}\approx 1/2$, the electroweak
potential ($\sqrt{2}{\rm \,}G_{{\rm F}}\,n_{e}$) is 
\begin{equation}
V_{W}=0.76\ Y_{e}{\frac{{\tilde{M}}}{10^{31}{\rm g}}\,}r{_{10}^{-3}}\times
10^{-12}\,\,{\rm eV}\;,  \label{VW}
\end{equation}
where $r_{10}=r/10^{10}\,{\rm cm}$. This is to be compared with 
\begin{equation}
A_{e}=1.48\ \frac{G_{e}}{G_{{\rm F}}}\frac{-\dot{L}_{e}}{10^{51}\,{\rm %
ergs/s/MeV}}\,r_{10}^{-2}\times {}10^{-12}\,\,{\rm eV\ ,}  \label{Ae2}
\end{equation}
where $G_{F}=11.66\,{\rm TeV}^{-2}$ is the Fermi constant and $%
G_{e}=1/\Omega _{e}^{2}$. 
Clearly, if the neutrino luminosities are
sufficiently high say, $10^{52}\,{\rm ergs/s}$ for $10{\rm \ MeV}$
neutrinos, and the scale of lepton symmetry breaking is around or below the
Fermi scale, the Majoron potential $A_{e}$ becomes competitive with $V_{W}$
at radii where the resonance occurs for $\Delta m^{2}$ values interesting
for solar neutrinos ($10^{-5}-10^{-4}\,{\rm eV}^{2}$). 
More generally, at
large enough distances the Majoron potentials decay as the inverse square
radius and necessarily dominate over the local interactions.

\begin{figure}[t]
\centering
\epsfig{file=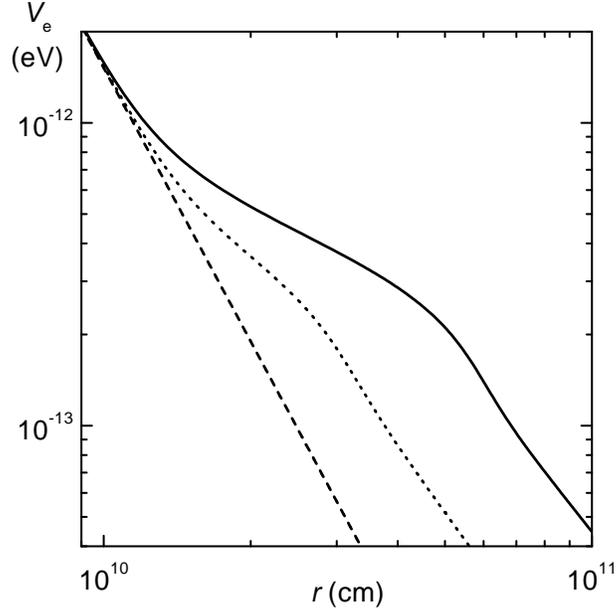,width=80mm}
\vspace{15pt}
\caption{Total potential $V_{e}=V_{\nu _{e}}-V_{\nu _{\mu }}$\ as a function
of the radius. The dashed curve stands for the SM potential, the dotted and
bold curves for the Majoron case with the same parameters as for the
homologous curves of Fig. 1.}
\end{figure}

Let us examine the $\nu _{e}\leftrightarrow \nu _{\mu }$\ oscillations with
the mixing parameters of the non-adiabatic solar neutrino solution (for a
recent update see \cite{hata97}), choosing in particular the values $\Delta
m^{2}=7\times 10^{-6}\,{\rm eV}^{2}$, $\Delta m^{2}\sin ^{2}2\theta =$\ $%
4\times 10^{-8}\,{\rm eV}^{2}$. In a supernova the resonance is
non-adiabatic as well and, as Eq.\ (\ref{Pc}) indicates, the survival
probability $P_{c}$ increases with the $\nu $ energy. This was studied in
detail in the framework of the SM \cite{mina88}. In Fig. 1, $P_{c}$ in the
Landau-Zener approximation is plotted against the energy. The dashed curve
holds for the SM potential with a constant $\tilde{M}=4\times 10^{31}{\rm g}$%
. It is manifest the aggravation of the non-adiabaticity with the energy.

To study the Majoron case one has to specify the energy spectra and
luminosities. I used Fermi-Dirac distributions with the following values of
temperature and chemical potential \cite{burr90}: for $\nu _{e}$, $T=2.4{\rm %
\ MeV}$ and$\ \mu =3.2\,T$; for $\nu _{\mu }$, $T=5.1{\rm \ MeV},\ \mu
=4.1\,T$. This gives average energies of $10$ and $23{\rm \ MeV}$
respectively. The luminosity intensities are in turn $10^{52}\,{\rm ergs/s}$
for $\nu _{e}$ and $7\times 10^{51}\,{\rm ergs/s}$ for $\nu _{\mu }$ which
amount to particle emission rates of $10^{51}$ and $3\times 10^{50}\,{\rm %
ergs/s/MeV}$ respectively.
 Because the $e$-neutrinos are more numerous, the $%
\nu _{e}\leftrightarrow \nu _{\mu }$\ oscillations produce a net destruction
of $L_{e}$-number and a positive Majoron potential $A_{e}$. The dynamics is
the following: the less energetic $\nu _{e}$\ oscillate to $\nu _{\mu }$\ at
the smallest radii; this conversion produces a positive $A_{e}(r)$\ which
attenuates the fall of the total potential $V_{e}=V_{W}+A_{e}$ with the
radius; as a consequence, the adiabaticity improves at larger $r$ and the
most energetic neutrinos change flavor with higher probabilities.

\begin{figure}[t]
\centering
\epsfig{file=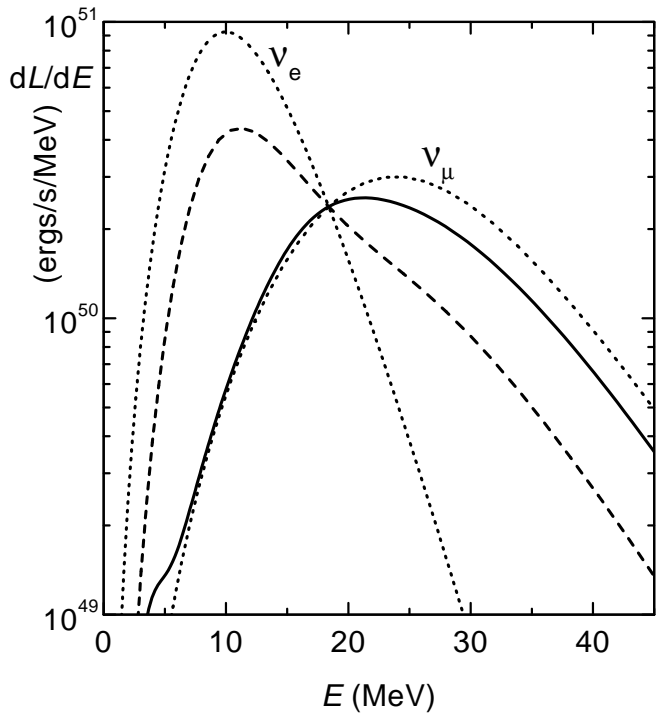,width=3.375in}
\vspace{15pt}
\caption{ The dotted curves are the assumed luminosity distributions for $%
\nu _{e}$ and\ $\nu _{\mu }$ as emitted from the neutrinospheres. The dashed
curve represents the $\nu _{e}$ luminosity after electroweak neutrino
oscillations whereas in the bold curve a $\xi _{e}$ field exists with $%
G_{e}=4\,G_{{\rm F}}$.}
\end{figure}

Figure 2 shows the total potential $V_{e}$\ as a function of the radius. The
dashed curve stands for $V_{W}$, the SM potential, with $\tilde{M}=4\times
10^{31}\,{\rm g}$\ and $Y_{e}=1/2$. In the dotted and bold curves the
Nambu-Goldstone field associated to $L_{e}$ symmetry breaking\ operates with
a constant $G_{e}=G_{{\rm F}}$ and $G_{e}=4\,G_{{\rm F}}$, \ respectively.
The potential falls more slowly as the field $\xi _{e}$ grows. The level
crossing probability is plotted in Fig.\ 1 for both cases (dotted and bold
curves) and the effect is clear: the stronger the Majoron field, the more
efficient is the flavor conversion. It is worth to mention that if the $\mu $%
-neutrinos were more numerous than the $e$-neutrinos the effect would be the
opposite because the Majoron potential would be negative ($\dot{L}_{e}>0$).
That is actually reflected in the rapid rise of $P_{c}$\ at the $\nu _{\mu }$%
\ energy band around $20{\rm \ MeV}$.

Figure 3 shows the implications for the outgoing $\nu _{e}$\ energy spectrum.
The dotted curves are the assumed luminosity distributions for the emitted $%
\nu _{e}$s and\ $\nu _{\mu }$s. The dashed curve represents the luminosity
of the $e$-neutrinos that come out of the star after standard MSW
oscillations and the bold curve is the same but with a Majoron field ($%
G_{e}=4\,G_{{\rm F}}$). The improvement of adiabaticity makes more $\nu
_{\mu }$s\ to convert into $\nu _{e}$\ and less $\nu _{e}$\ to survive, and
because the $\mu $-neutrinos are more energetic, the outgoing $\nu _{e}$\
spectrum is harder than if there was no Majoron field. The average energy of
the outgoing $e$-neutrinos\ is $13{\rm \ MeV}$ if $G_{e}=0$ but rises to $17%
{\rm \ MeV}$ if $G_{e}=G_{{\rm F}}$ and $21{\rm \ MeV}$\ if $G_{e}=4\,G_{%
{\rm F}}$.

\section{Conclusions and discussion}

To summarize, if the explanation of the solar neutrino deficit is the MSW
non-adiabatic oscillation $\nu _{e}\rightarrow \nu _{\mu }\ $(or $\nu
_{e}\rightarrow \nu _{\tau }$) then, the standard model of electroweak
interactions predicts that in a supernova the $\nu _{e}\leftrightarrow \nu
_{\mu }$\ transitions are also non-adiabatic. It means that, to a large
extent, the $e$-neutrinos preserve their lower energy spectrum, unless $\nu
_{e}$\ also mixes to another flavor with a too high or too low $\Delta m^{2}$%
\ to show up in solar neutrinos. If however, $L_{e}$\ is a spontaneously
broken quantum number, the associated Nambu-Goldstone boson, $\xi _{e}$,
will acquire a classic field configuration which may be strong enough to
produce a back reaction with the net effect of improving the adiabaticity of
the $\nu _{e}\leftrightarrow \nu _{\mu }$\ transitions. The final result is
a $\nu _{e}$\ energy spectrum harder than expected.

In 1987, the existing detectors were only able to detect electron
anti-neutrinos but the now operating Super-Kamiokande and SNO experiments
will be capable of detecting supernova $\nu _{e}$\ events. The analysis of
the energy distribution can in principle reveal or put limits on that kind
of effect.

The scenarios of neutrino mixing change considerably if one considers the
evidences from atmospheric and terrestrial neutrino experiments (for a
review see \cite{cald98}). The atmospheric neutrino anomaly and the zenith
angle dependence observed by Super-Kamiokande can be explained by $\nu _{\mu
}\rightarrow \nu _{\tau }$ oscillations, best fit \cite{suzu98} $\Delta
m^{2}=5\times 10^{-3}\,{\rm eV}^{2}$, $\sin ^{2}2\theta =1$. The alternative 
$\nu _{\mu }\rightarrow \nu _{e}$ is excluded by the CHOOZ limits \cite
{CHOOZ97}. This can still accommodate $\nu _{e}\rightarrow \nu _{\mu }$ or $%
\nu _{e}\rightarrow \nu _{\tau }$ as solar neutrino solutions but that is no
longer true if one takes in consideration the evidence from the
Liquid Scintillation Neutrino Detector (LSND)
 for $\bar{\nu}_{\mu }\rightarrow \bar{\nu}_{e}$\ \cite{LSND96} 
and \ $\nu _{\mu
}\rightarrow \nu _{e}$ \cite{LSND97} oscillations. The very different $%
\Delta m^{2}$ scales involved in LSND ($\Delta m_{e\mu }^{2}>0.2\,{\rm eV}%
^{2}$), atmospheric and solar neutrinos call for a fourth flavor - a sterile
neutrino. In that picture solar $\nu _{e}$s oscillate into the sterile $\nu
_{s}$.

We now examine the consequences for supernovae neutrinos always assuming the
non-adiabatic solar neutrino solution, ignoring for definiteness the
possible mixing between $\nu _{s}$ and $\nu _{\mu }$ or $\nu _{\tau }$. The
oscillation pattern is the following: 1) MSW conversion of $\nu
_{e}\rightarrow \nu _{\mu }$ with LSND $\Delta m^{2}$; 2) sequential
oscillation $\nu _{\mu }\rightarrow \nu _{e}\rightarrow \nu _{s}$, the first
a LSND transition, the second a solar $\nu $ process. The outcome is a hard
spectrum for $\nu _{e}$ depleted by $\nu _{e}\rightarrow \nu _{s}$, but only
partially because of the non-adiabaticity of this transition. If
alternatively, a NG field $\xi _{e}$ exists (created by $\nu _{e}\rightarrow
\nu _{\mu }$, $\nu _{\mu }\rightarrow \nu _{e}$ and $\nu _{e}\rightarrow \nu
_{s}$), it improves the adiabaticity of $\nu _{e}\rightarrow \nu _{s}$
causing a $\nu _{e}$ depletion stronger than predicted by SM interactions.

If one repeats the analysis with other scenarios of $\nu $ mixing the
effects will be different in detail but with one thing in common: the
signature of NG fields is a {\em surprise} {\em i.e.}, an oscillation
pattern not consistent with the $\nu $ mixing derived from solar,
atmospheric and terrestrial $\nu $ experiments. It should be kept in mind
that the situation turns more complex and rich if there is mixing between
the three NG bosons associated with the three lepton flavors, a very natural
feature if they are all spontaneously broken. This was explored in \cite
{bent98}.

A point that cannot be overstressed is that the Nambu-Goldstone fields are
proportional to the rate of charge violation processes and therefore to the
very reaction rates. In the case of neutrino oscillation this manifests as a
strong dependence of the Majoron fields on the magnitudes of the neutrino
luminosities. In the numeric simulation I chose $10^{52}\,{\rm ergs/s}$ for $%
\nu _{e}$ and $7\times 10^{51}\,{\rm ergs/s}$ for $\nu _{\mu }$,\ values
produced and even exceeded in the about half a second that lasts between the
neutronisation $\nu _{e}$ burst and the supernova explosion \cite
{burr92,mayl87,burr90}. The neutrino luminosities decay afterwards in a time
scale of 1 second, or rather 4 seconds \cite{burr90,blud88},\ as indicated
by SN 1987A events \cite{hira87}. The highest luminosity happens during the
first $\nu _{e}$ burst - above $10^{53}\,{\rm ergs/s}$ in the peak \cite
{burr92,mayl87,burr90} - and the Majoron field may be even stronger then.
However, the time scale of the rise and fall of the $\nu _{e}$ signal is
about $5\,{\rm ms}$, too short to authorize a stationary approximation in
the calculation of $\xi _{e}$. In fact, distances of the order of $10^{10}\,%
{\rm cm}$ in such a period of time are beyond the light cone and a special
study is required.

The effects of the Majoron fields on the neutrino spectra, if any, will be
observed in a shorter or longer interval of time depending on the actual
scale of lepton number symmetry breaking. The observation of such a
correlation with the flux magnitudes, by itself a signature of the NG
fields, would thus provide a measurement of the scale of spontaneous
symmetry breaking.

\strut

\acknowledgements
This work was supported in part by the project ESO/P/PRO/1127/96.


\strut

\strut

\end{document}